\documentclass[fleqn,usenatbib]{mnras}

\usepackage{ae,aecompl}

\usepackage{graphicx}	
\usepackage{amsmath}	
\usepackage{amssymb}	

\usepackage{txfonts} 

\usepackage{journals}

\usepackage[none]{hyphenat}

\title[Reconstructing the velocity dispersion profiles]{Reconstructing the velocity dispersion profiles from the 
line-of-sight kinematic data in disc galaxies}
\author[A. A. Marchuk and N. Y. Sotnikova]
{A. A. Marchuk$^{1}$\thanks{E-mail:
a.marchuk@spbu.ru (AAM)} and N. Y. Sotnikova$^{1,2}$\\
$^{1}$St. Petersburg State University,
Universitetskij pr.~28, 198504 St. Petersburg, Stary Peterhof, Russia\\
$^{2}$Isaac Newton Institute of Chile, St. Petersburg Branch, Russia\\
}

\date{Accepted XXX. Received YYY; in original form ZZZ}

\pubyear{2016}

\begin{document}
\label{firstpage}
\pagerange{\pageref{firstpage}--\pageref{lastpage}}
\maketitle

\begin{abstract}
We present a modification of the method for reconstructing the 
stellar velocity ellipsoid (SVE) in disc galaxies. Our version 
does not need any parametrization of the velocity dispersion 
profiles and uses only one assumption that the ratio 
$\sigma_z/\sigma_R$ remains constant along the profile or 
along several pieces of the profile. The method was tested 
on two galaxies from the sample of other authors and for the 
first time was applied to three lenticular galaxies 
NGC~1167, NGC~3245 and NGC~4150 as well as to one Sab galaxy NGC~338. 
We found that for galaxies with a high inclination ($i >55-60\degr$)
it is difficult or rather impossible to extract the information 
about SVE while for galaxies at an intermediate inclination 
the procedure of extracting is successful. For NGC~1167 we 
managed to reconstruct SVE, provided that the value 
of $\sigma_z/\sigma_R$ is piecewise constant. We found 
$\sigma_z/\sigma_R=0.7$ for the inner parts of the disc and 
$\sigma_z/\sigma_R=0.3$ for the outskirts. We also obtained a 
rigid constrain on the value of the radial velocity dispersion 
$\sigma_R$ for highly inclined galaxies and tested the result using 
the asymmetric drift equation, provided that the gas rotation curve 
is available.
\end{abstract}


\section{Introduction}

One of the most challenging problems of galactic dynamics is 
reconstructing the distribution function (DF) of a certain galaxy from 
its limited observational data including surface photometry and 
long-slit or two-dimensional field spectroscopy. Various techniques 
have been invented last several years in order to solve this problem.

The most developed scheme is based on Schwarzschild linear 
programming method \citep{Schwarzschild1979}. It uses a library 
of orbits in a given potential to reproduce the structure and 
kinematics of a stellar system, mainly triaxial models 
(see, e.g. \citealp{Capuzzo-Dolcetta_etal2007,vandenBosch_etal2008} 
and references therein). There is a particle-based extension of 
the Schwarzscild' method (made-to-measure method). It works by 
adjusting individual particle weights as the model evolves, until 
the $N$-particle system reproduces a set of target constraints 
(see e.g. 
\citealp{Syer_Tremaine1996,deLorenzi_etal2007,deLorenzi_etal2008}). 
A new iterative method for constructing equilibrium phase models of 
stellar systems \citep{Rodionov_Sotnikova2006,Rodionov_etal2009} 
has been tested on galaxies with imitated observational data and has
shown its flexibility \citep{Rodionov_etal2009}.

All these algorithms need a detailed mass model of a target galaxy, 
so nearly all dynamical models have been applied for one-component 
system, mainly for elliptical galaxies. There are still no 
reliable complete phase models of observed spiral galaxies that are 
multicomponent systems with a substantial mass contribution of 
invisible matter -- dark haloes. One obvious exception is the 
Milky Way with a detailed phase model~\citep{Bovy_Rix2013}.

In many cases we do not need a DF to judge about dynamic status of 
a stellar disc since it is sufficient to know the velocity dispersion 
profiles, mainly $\sigma_{R}(R)$ and $\sigma_{z}(R)$. The radial 
velocity dispersion profile $\sigma_{R}$ gives 
$Q = \sigma_{R}/\sigma_{R}^\mathrm{cr}$ \citep{Toomre1964}, 
where $\sigma_{R}^\mathrm{cr} = 3.36 \, G \, \Sigma_\mathrm{s}/ \kappa$ 
for an infinitely thin stellar disc, $\kappa$ is the epicyclic 
frequency and $\Sigma_\mathrm{s}$ is the disc surface density. 
The $Q$-parameter shows the level of disc heating and stability 
against perturbations in a disc plane. The ratio 
$\sigma_{z}/\sigma_{R}$ describes the shape of the stellar velocity 
ellipsoid (SVE) and tells about the relaxation in a disc and 
its dynamical history.

The $\sigma_{z}/\sigma_{R}$ ratio for stars in the solar neighbourhood 
is $0.53\pm0.07$ \citep{Dehnen_Binney1998}. Until recently, there were no data available for external galaxies 
about the $\sigma_{z}/\sigma_{R}$ ratio. Even for the nearby galaxy 
M31 individual stellar measurements give only the line-of-sight 
profile of the velocity dispersion~\citep{Dorman_etal2015}.
Observations of external galaxies 
have been concentrated on systems that are close either to face-on 
or to edge-on cases
\citep[e.g.][]{vanderKruit_Freeman1986,Bottema1993,Kregel_etal2005}. 
Face-on systems provide direct information only about one component 
of random velocities -- $\sigma_{z}(R)$. For edge-on systems we do not obtain $\sigma_{R}(R)$ or 
$\sigma_{\varphi}(R)$ directly because of the integration of the 
DF along the line-of-sight. The profile 
$\sigma_{R}(R)$ can be obtained 
implicitly via the asymmetric drift equation that describes 
the equilibrium in a plane of a rotating stellar disc. 
The procedure needs the usage of the gas and stellar velocity curves. 
A major concern is that the integration along the line-of-sight can have a
dramatic effect on the derived rotation curve. 
This effect leads to a marked discrepancy between the rotation curve 
and the circular velocity curve especially in the inner part of a 
galaxy \citep{Zasov_Khoperskov2003,Stepanova_Volkov2013}. 
As a result the scheme needs the restoration of the rotation curve 
\citep{DiTeodoro_Fraternali2015} and more or less robust 
parametrization of $\sigma_{R}(R)$ \citep{Kregel_etal2005}.

All three moments of the random velocity 
distribution in a stellar disc can be found for a galaxy at an 
intermediate inclination. Long-slit data from a variety of position 
angles and large-area integral field units provide information about 
the line-of-sight velocity dispersion $\sigma_\mathrm{los}$ at different 
points in a galaxy. All three components 
$\sigma_{R}$, $\sigma_{\varphi}$ and $\sigma_{z}$ contribute 
to $\sigma_\mathrm{los}$ depending on the azimuthal angle and 
inclination. There are two ways to derive these components. 
One way is to construct a set of maximum-mass marginally stable 
discs ($N$-body models). Then the line-of-sight velocity dispersion 
profiles can be constructed and compared with observational data 
to choose the most reliable model 
\citep{Bottema_Geritsen1997,Zasov_etal2008,Zasov_etal2012}. 
Another way is to pick out $\sigma_{R}$, $\sigma_{\varphi}$ 
and $\sigma_{z}$ directly from $\sigma_\mathrm{los}$ along the major 
and minor axes, provided the epicycle approximation that connects 
$\sigma_{R}$, $\sigma_{\varphi}$ and the rotational velocity of 
stars $\bar{v}_{\varphi}$ 
\citep{Gerssen_etal1997,Gerssen_etal2000,Shapiro_etal2003,Gerssen_Shapiro2012,Verheijen_etal2004,Noordermeer_etal2008}.

To derive the velocity dispersion profiles directly exponential 
approximations for both radial and vertical velocity 
dispersion components are usually assumed. The most conventional 
ansatz of a such parametric approach is 
$\sigma_{z}^{2}(R) \propto \Sigma_\mathrm{s}(R)$ 
(the assumption is valid for an isothermal layer with a constant 
thickness, \citealp{Spitzer1942}) 
and $\sigma_{z}/\sigma_{R} = \mathrm{const}$ 
throughout a disc \citep{Verheijen_etal2004}. There are numerical 
bases that stellar discs are vertically isothermal 
\citep[e.g.][]{Rodionov_Sotnikova2006}. For an exponential disc 
with a scalelength $h$ the assumption 
$\sigma_{z}^{2}(R) \propto \Sigma_\mathrm{s}(R)$ results in 
$\sigma_{z}(R) \propto \exp(-R/2h)$ and this agrees fairy well with 
observational data \citep{vanderKruit_Searle1981}. The assumption $\sigma_{z}/\sigma_{R} = \mathrm{const}$ leads 
to the exponential profile for $\sigma_{R}$ with a scalelength $2h$ 
(see, e.g. \citealp{Martinsson_etal2013}).
The above assumption has no clear evidences and is not supported 
by $N$-body simulations (see, e.g. \citealp{Minchev_etal2012}) but 
is commonly accepted in theoretical analysis.

\cite{Gerssen_etal1997,Gerssen_etal2000}, \cite{Shapiro_etal2003},
\cite{Gerssen_Shapiro2012} 
used a more free parametrization of the velocity dispersion 
profiles. They let the scale length for $\sigma_{R}(R)$ and 
$\sigma_{z}(R)$ be not equal to $2h$ and obtained that it was a 
factor 3--5 larger than the disc scale length $h$. The velocity 
dispersion profiles derived by \citet{Noordermeer_etal2008} 
from spectroscopic data along the major and minor axes got also very 
shallow. These results contradict with the approximation of 
isothermal layers and force to revise the procedure of extracting 
the velocity dispersion profiles from the line-of-sight kinematics.

Instead of using any parametrization for the velocity dispersion 
profiles like 
\cite{Gerssen_etal1997,Gerssen_etal2000}, \cite{Shapiro_etal2003},
\cite{Gerssen_Shapiro2012}, in
\citet{Silchenko_etal2011} authors followed another approach. They used an 
equation containing velocity dispersions and difference between 
the stellar rotation curve and the local circular speed (the so-called 
asymmetric drift equation, see equation~\ref{eq:AsDr}). If one has data 
for both gas and stellar rotation curves then this equation might 
be solved iteratively. \citet{Silchenko_etal2011} applied this 
approach to the early-type disc galaxy NGC~7217. Retrieved 
independently from the major and minor line-of-sight data 
two profiles of $\sigma_{z}(R)$ are in good agreement
and derived $\sigma_{R}(R)$ profile looks reasonable too. 
However, it is quite difficult to make conclusions about method's 
quality based on one example.

\citet{Noordermeer_etal2008} also avoid exponential parametrization 
and solve a system of equations directly with one additional condition 
$\sigma_{z}=\sigma_{\varphi}$. The authors also supposed the ratio 
$\sigma_{z}/\sigma_{R}$ to be constant and varied it in a range 
from 0 to 0.82. \cite{Noordermeer_etal2008} used this ratio for 
reconstructing the gas rotation curve from the asymmetric drift 
equation. Despite the good result, authors noted that their 
assumption had no real physical basis.

We revise the procedure of extracting the SVE from the 
line-of-sight kinematics, consistently rejecting some 
assumptions made in previous works.

The outline of this paper is as follows.

In Section 2, there is a list of four spiral galaxies with spectral 
data for the major and minor axes of a disc.
In Section 3, a general scheme for SVE reconstruction is given. 
In Section 4, we describe how the procedure of SVE 
extracting was modified. 
In Section 5, some tests and main results are presented.
Section 6 discusses the results and related questions, and 
Section 7 contains conclusions.

\section{Data}

Our dataset consists of three lenticular S0 (NGC~1167, NGC~3245 and 
NGC~4150) and one Sab galaxies -- NGC~338 
(see Table~\ref{table:main_parameters}). 
The data for NGC~1167 and NGC~4150 have been provided by 
Alexei Moiseev from the Special Astrophysical Observatory of the
Russian Academy of Sciences and the data for NGC~338 and NGC~3245 
have been received from Ivan Katkov (Sternberg Astronomical
Institute of the Moscow State University. Other three galaxies 
from these papers were not included because we select only galaxies 
with well-defined data across minor axis. The data were obtained with 
the SCORPIO \citep{Afanasiev_2005} focal reducer in the long-slit mode (with the slit size 
of $6\arcmin \times 1\arcsec$) mounted at the primary focus of the 
6-m telescope of the Special Astrophysical Observatory. Observations 
and data reduction are described in detail in 
\citet{Zasov_etal2008} for NGC~1167 and NGC~4150 and in 
\citet{Zasov_etal2012} for NGC~338 and NGC~3245.

Observations were made in the wavelength range 4800--5540~\AA,
which contains numerous absorption lines of the old
stellar populations in the galaxies. 
The spectral resolution was 2.2~\AA\, for NGC~1167 and NGC~4150 and 
2.6~\AA\, for NGC~338 and NGC~3245.

The classical cross-correlation method, similar to employed in 
\citet{Moiseev2001} was used to calculate the radial velocities 
and the stellar velocity dispersions for NGC~1167 and NGC~4150. 
The ULySS software package was used for the pixel by pixel 
approximation of the observed spectra by the model spectra of 
stellar populations for NGC~338 and NGC~3245.
For this purpose, the high-resolution PEGASE.HR \citep{Borgne_2004} models with SSP 
(Simple Stellar Population) were applied to binarized spectra for all 
galaxies. Adaptive binning of the spectra along the slit was used 
with the signal-to-noise ratio in each element not less than 
20-50.
At every spectra bin fitting produces 
radial velocity $V_r$ and the velocity dispersion $\sigma$.

Below we listed all the galaxies with their peculiarities. This is a 
summary of the analysis done 
by~\citet{Zasov_etal2008,Zasov_etal2012}.

${\bf NGC~338}$ is a type Sa or Sab galaxy. The images from Sloan Digital Sky Survey  
show blurred spiral structures in the disc. 
The bulge is anomalously bright and extended according 
to~\citet{Noordermeer_vanderHulst2007} but 
this conclusion may be due to the inaccurate photometric 
decomposition of brightness. 
The galaxy contains a large (for lenticular galaxies) 
amount of HI gas.

NGC~338 is a fast rotating galaxy with maximum circular velocity near 
280-320~km/s. The velocity profiles were obtained from absorption 
lines and emissions in $\mathrm{H}_{\beta}$, [NI] and [OIII]. 
Both gas and stellar rotation curves are measured up to $50\arcsec$ 
and have velocity uncertainties not greater than~10~km/s. 
The rotation curve reaches the plateau at $r \approx$~4~kpc 
($12.5\arcsec$), but velocities at distances from the centre larger 
than 4~kpc become asymmetrical. Such an asymmetry in the velocity 
field may be due to a small merging event. Gas velocities 
exceed stellar ones noticeably. This fact can be explained 
by a large velocity dispersion. 
{The inclination and the position 
angle are poorly constrained from kinematics and were derived from 
the optical isophotal analysis~\citep{Noordermeer_vanderHulst2007b}.}
The line-of-sight velocity dispersion value along the major axis 
(PA = $108\degr$) is high inside the bulge-dominated region 
($r < 15\arcsec$) and then decreases rapidly until it becomes constant at 
$20\arcsec$. The velocity dispersion along the minor axis 
(PA = $17\degr$) has uncertainties about 15-20~km/s.

${\bf NGC~1167}$ does not belong to any group and 
does not have satellites with comparable luminosity. The bulge has 
a steep photometric profile (the S\'ersic index $n \approx 3$ 
according to \citealp{Zasov_etal2008}) and dominates over the disc 
up to $15-20\arcsec$. The total disc luminosity is three 
times greater than the bulge luminosity. 
According to \citet{Noordermeer_etal2005}, the galaxy 
contains a significant amount of HI (the total mass 
$\approx 1.7 \times 10^{10} M_{\sun}$) that is spread over a large area. 
The average HI surface density remains below the critical value 
for the gravitational instability of gaseous layer. 

Observations of neutral hydrogen show a rotation curve extended over 
the 10 exponential disc scales. The maximum circular velocity is 
very large and is not far from 400~km/s (at $100\arcsec$). 
Stellar rotation velocities are measured up to $r = 50\arcsec$ 
and are similar to gas values and uncertainties (10--15~km/s). 
Line-of-sight velocity dispersion profiles along the major and minor 
axes have average uncertainties around 20~km/s. They also have the same slope 
and their central values are close. All these factors do not allow us 
to separate the profiles from each other in central parts with sufficient confidence.

The inclination and the position angle for this galaxy were derived 
from the analysis of the velocity 
field~\citep{Noordermeer_vanderHulst2007b}.

${\bf NGC~3245}$ is an isolated galaxy with the inner and outer 
discs. The border between the bright inner and outer discs 
is very sharp and falls on $r \approx 15-20\arcsec$. 
No spirals, HII areas and active star formation regions were found. 
Galaxy has a compact bulge with effective radius 
approximately equals $5\arcsec$.

\begin{figure*}
\includegraphics[width=2\columnwidth]{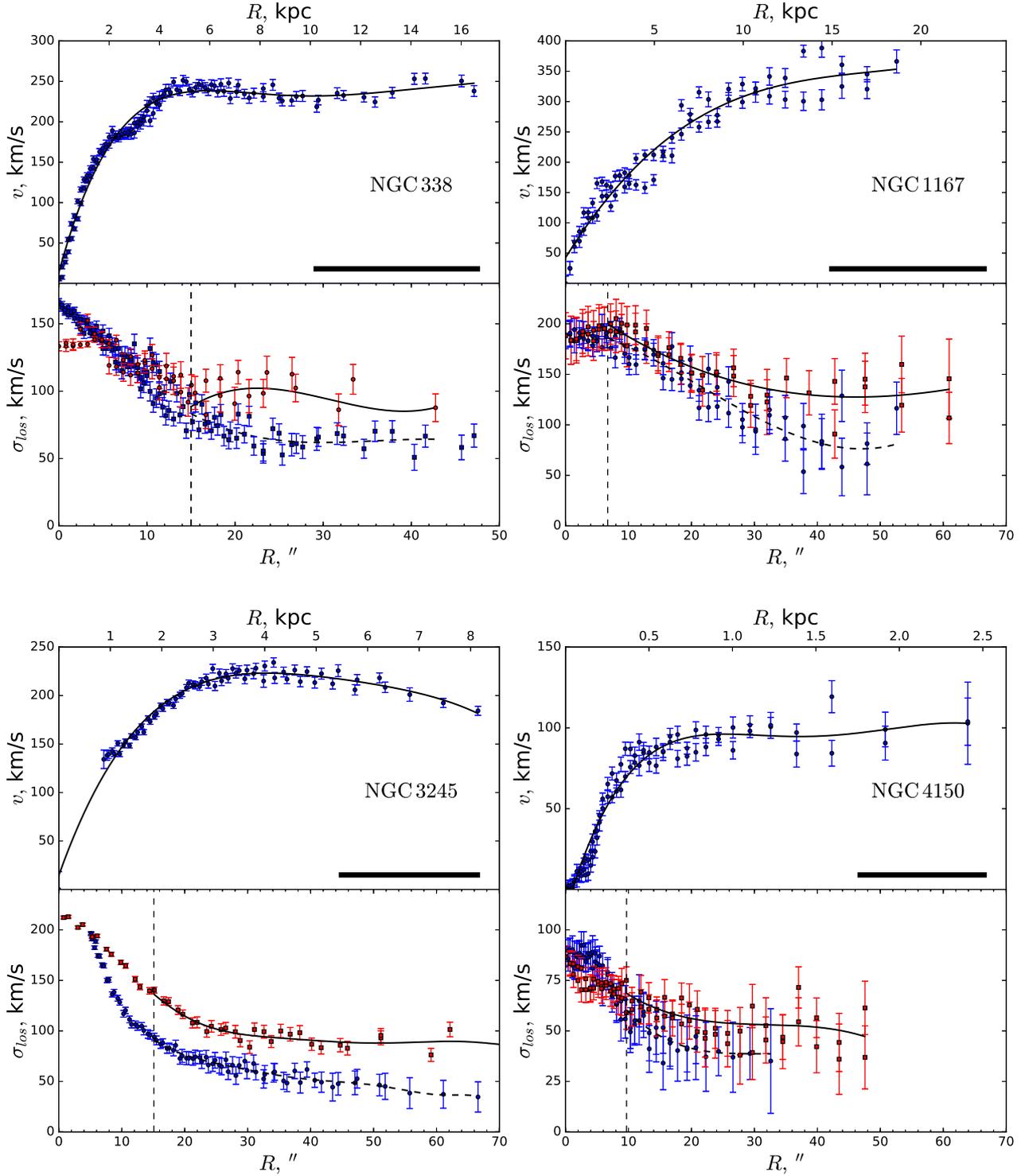}
\caption{The radial velocities and the velocity dispersions for 
the examined galaxies. Each galaxy is represented by two plots. 
The top plot shows the observational points for the stellar 
rotation curve, their approximation by splines is shown by the solid 
line. In the lower right corner a line segment is plotted. Its 
length corresponds to the disc scalelength 
(see Table~\ref{table:main_parameters}). 
The bottom plot shows the observational data for the 
line-of-sight stellar velocity dispersions along the major axis 
$\sigma_\mathrm{los,maj}$ (filled blue circles) and along the minor 
axis $\sigma_\mathrm{los,min}$ (filled red squares). The solid and 
dashed lines are the best approximation of the profiles along 
the minor and major axes, respectively. 
The vertical dashed line separates the region $R < r_\mathrm{e,b}$ 
where bulge dominates.}
\label{fig:obs_data} 
\end{figure*}

The rotation curve and the velocity dispersion profiles are 
symmetrical, but the galaxy has some dynamical peculiarities. 
First, it may be possible that the gaseous disc is inclined to the 
stellar disc, because gas velocities along the minor axis have 
non-zero gradient. 
Second, the stellar rotation curve has a maximum at 
$r \approx 2.5$~kpc (around $30\arcsec$) and slowly decreases after. 
Also galaxy has huge central velocity dispersion values 
(similar to NGC~1167, see Figure~\ref{fig:obs_data}). 
The line-of-sight velocity dispersion profiles have small 
uncertainties but they diverge in the centre 
($|\sigma_\mathrm{los}^\mathrm{maj} - \sigma_\mathrm{los}^\mathrm{min}| 
\approx 20$~km/s). Perhaps this indicates that the slits were shifted 
relative to the centre.

${\bf NGC~4150}$ is a low-luminosity galaxy and has the exponential 
photometric profile. There is no data obtained for the gas motions. 
The stellar rotation curve remains nearly flat for $r > 20\arcsec$ and 
has maximum velocity $\approx$~100~km/s. The velocity dispersion 
data are very noisy and have the biggest uncertainties in the whole dataset. Due to these reasons, the fitting of the line-of-sight velocity dispersion profiles becomes complicated. For NGC~3245 and NGC~4150 the photometrically determined orientation angles of the disc were used.

\begin{table*}
\caption{The main parameters of the examined galaxies.}
\label{table:main_parameters}
\begin{tabular}{ c c c c c c c c }
\hline
Galaxy & Type & Inclination $i$, degr & Distance, Mpc & scale, kpc/arcsec & 
$M_\mathrm{B}$ & $r_\mathrm{e,b}$, arcsec & $h$, arcsec\\
\hline
NGC~338 & Sab & $64^{\rm I}$ & 62 & 0.292 & -21.48 & $15.0^{\rm a}$ & $18.3^{\rm a}$\\ 
NGC~1167 & S0 & $36^{\rm I}$ & 66 & 0.310 & -21.73 & $6.7^{\rm a}$ & $24.2^{\rm a}$\\ 
NGC~3245 & S0 & $62^{\rm II}$ & 22.3 & 0.107 & -20.17 & $15.1^{\rm b}$ & $21.6^{\rm b}$\\ 
NGC~4150 & S0 & $56^{\rm II}$ & 6.7 & 0.033 & -18.51 & $9.5^{\rm c}$ & $19.7^{\rm c}$\\
\hline 
\end{tabular}
\\
Morphological type (2) was taken from \protect\citet{Zasov_etal2008} 
and \protect\citet{Zasov_etal2012}. 
Inclination angle (3) references are 
(I) \protect\citet{Noordermeer_vanderHulst2007b},
(II) LEDA database (Lyon/MeudonExtragalactic Database). 
The distance to the galaxy (4) and the scale (5) were found in the 
NASA / IPAC Extragalactic Database (NED). 
Data for the Hubble constant $H_\mathrm{o}$ = 73.00~km/s/Mpc, 
$\Omega_\mathrm{matter}$ = 0.27, $\Omega_\mathrm{vacuum}$ = 0.73. 
The absolute magnitude in the $B$ band (6) was taken from the LEDA 
database. Photometric decomposition data, namely the effective 
bulge radius (7) and the exponential disc scale (8), were taken from 
(a) \protect\citet{Noordermeer_vanderHulst2007} in the $R$ band, 
(b) \protect\citet{Fisher_Drory2010} and 
(c) \protect\citet{Fabricius_etal2012} in 3.6~$\mu$m.
\end{table*}

\section{Stellar velocity dispersions}

In cylindrical polar coordinates 
$(R, \, \varphi, \, z)$ the line-of-sight velocity dispersion along 
a slit relates to the SVE components of a thin disc 
$\sigma_{R}$, $\sigma_{\varphi}$, $\sigma_{z}$ as
\begin{equation}
\sigma_\mathrm{los,\phi}^{2} = 
[ \sigma_{R}^{2}\,\sin^{2}\phi + 
\sigma_{\varphi}^{2}\,\cos^{2}\phi ] \, 
\sin^{2}i + \sigma_{z}^{2} \, \cos^{2}i \, ,
\label{eq:los_comm}
\end{equation}
where $\phi$ is the angle between the slit and the major axis of 
the projected stellar disc (the position angle), $i$ is the galaxy 
inclination angle. For the observed line-of-sight velocity dispersions along minor and major axes we can write:
\begin{equation}
\begin{array}{rcl}
\sigma_\mathrm{los,min}^{2} & = 
& \sigma_{R}^{2}\,\sin^{2}i + \sigma_{z}^{2} \, \cos^{2}i \, , \\
\sigma_\mathrm{los,maj}^{2} & = 
& \sigma_{\varphi}^{2} \, \sin^{2} i + \sigma_{z}^{2} \, \cos^{2}i \, .
\end{array}
\label{eq:slos}
\end{equation}

Equations~\eqref{eq:los_comm} and~\eqref{eq:slos} are valid for a thin 
disc. For discs at high inclination the assumption about 
a thin disc in these equations 
may breaks down as soon as the line-of-sight integration through the 
disc involves $\sigma_{R}$ from substantial regions through the disc 
but not from a given $R$. That is why the usage of the above 
equations for discs at high inclinations seems to be unreliable.

The system of two equations contains three unknowns $\sigma_{R}$, 
$\sigma_{\varphi}$ and $\sigma_{z}$. It may be closed by some additional 
dynamical relation, which is valid if the whole system is in 
equilibrium. One such relation connects the radial and 
the azimuthal velocity dispersion components with the mean azimuthal 
velocity of stars $\bar{v}_{\varphi}$ \citep{Binney_Tremaine2008}
\begin{equation}
f(R) = \frac{\sigma_{\varphi}^{2}(R)}{\sigma_{R}^{2}(R)} = 
\frac{1}{2}\left(1 + 
\frac{\partial \ln \bar{v}_{\varphi}}{\partial\ln R}\right) \, .
\label{eq:vphi}
\end{equation}

Due to the noise presented in the data it is difficult to obtain 
all three components of the velocity dispersion directly from three 
equations presented above \citep{Gerssen_etal1997} because the 
procedure includes subtraction of very close values 
$\sigma_\mathrm{los,min}^{2}$ and $\sigma_\mathrm{los,maj}^{2}$. 
A possible solution is to parametrize the profiles and to find 
the best-fitting parameters. Such an approach has been used in many papers 
\citep{Gerssen_etal1997,Gerssen_etal2000,Shapiro_etal2003,Gerssen_Shapiro2012}, 
but it depends on the selected parametrization, which is usually 
exponential. \citet{Noordermeer_etal2008} introduced less parametric 
approach using equation for asymmetric drift (hereafter AD) and 
non physical assumption that $\sigma_{\varphi} = \sigma_{z}$. 
Equation for AD shows difference between the mean stellar rotation velocity and the local circular speed~\mbox{\citep{Binney_Tremaine2008}}:
\begin{equation}
v_\mathrm{c}^{2}-\bar{v}_{\varphi}^{2} = \sigma_{R}^{2} 
\left(\frac{\sigma_{\varphi}^{2}}{\sigma_{R}^{2}} - 1 - 
\frac{\partial\ln\Sigma_\mathrm{s}}{\partial\ln R} - 
\frac{\partial\ln\sigma_{R}^{2}}{\partial\ln R}\right) \, ,
\label{eq:AsDr}
\end{equation}
where $\Sigma_\mathrm{s}$ is the stellar surface density. Notice that 
this form of the equation neglects the tilt term 
$d(\overline{v_{R}v_{z}})/dz$, which is assumed to be small compared 
to the other terms. The method by \citet{Noordermeer_etal2008} was 
adopted by \citet{Silchenko_etal2011} for the exponential velocity 
dispersion profiles and applied to NGC~7217 and its spectral data. 
\citet{Silchenko_etal2011} supposed to use the brightness 
for the old stellar population in the $I$ band instead of the surface density 
in equation~\eqref{eq:AsDr} and then found the $\sigma_{R}$ profile iteratively.

In this paper we present a new approximation of the line-of-sight velocity dispersion profiles without exponential 
fits. We adopt only one 
additional condition that the ratio $\sigma_{z}/\sigma_{R}$ is 
constant or piecewise constant throughout the disc. 
This statement is questionable as it is an oversimplification.

Using this assumption, we infer from the first equation 
in equations~\eqref{eq:slos} that $\sigma_\mathrm{los,min}$ is proportional 
to $\sigma_{R}$ with a proportionality factor 
$\sin^{2}i + \alpha^2\cos^{2}i$, where $\alpha$ hereafter means 
the ratio of vertical and radial velocity dispersions $\sigma_{z}/\sigma_{R}$. This leads to conclusion that for 
every radius $R$ the next equality is true: 
\begin{equation}
F(R) = \frac{\sigma_\mathrm{los,min}(R)} 
{\sigma_\mathrm{los,min}(R_0)} = 
\frac{\sigma_{R}(R)}{\sigma_{R}(R_0)} \, ,
\label{eq:f}
\end{equation}
for any $R_0$ in a suitable range. The selection of the $R_0$ value 
is described in the next section. We define 
$\sigma_{R, 0} = \sigma_{R}(R_0)$. From the proportion above and 
from~\eqref{eq:slos} we can derive the final set of equations with 
two unknown variables $\alpha$ and $\sigma_{R, 0}$:
\begin{equation}
\begin{array}{lcl}
\sigma_\mathrm{los,min}^{2} = 
\sigma_{R,0}^2\, F^2 \left(\sin^{2}i + 
\alpha^{2}\,\cos^{2}i\,\right) , \\
\sigma_\mathrm{los,maj}^{2} = 
\sigma_{R,0}^2\, F^2 \left(f\sin^{2}i + 
\alpha^{2}\,\cos^{2}i\,\right).
\end{array}
\label{eq:final}
\end{equation}
In the next sections we use equations~\eqref{eq:vphi} and~\eqref{eq:final} to 
derive SVE $(\sigma_{R}$, $\sigma_{z}$, $\sigma_{\varphi})$ 
and invoke the AD equation~\eqref{eq:AsDr} to verify radial dispersion values when it is possible.

\section{The method}
\label{subs:reconstr}

The stellar rotation curves were corrected for the mean velocity, 
folded around the centre and reduced for the inclination factor 
$\sin i$. The resulting curves were fitted by the best 
smoothing B-spline curve with $k = 3$ which is stay unchanged in further analysis. The same corrections and approximation have been applied to the gas data if they were available.

Both stellar velocity dispersion profiles along the major and minor 
axes were bended around the appropriate centre. After that 
$\sigma_\mathrm{los}^\mathrm{min}$ data were deprojected and 
reduced for the inclination factor $\cos^{-1}i$. Then, all data 
at radii less than effective bulge radius $r_\mathrm{e,b}$ were 
removed to avoid the bulge influence in the central region and 
contamination of profiles. Photometric scalelengths for the examined 
galaxies are listed in Table~\ref{table:main_parameters}. 
Both line-of-sight velocity dispersion profiles were fitted by smoothing B-spline 
with $k = 3$ in the range from $R_0$ to the last point of the data. 
In order to take into account different errors in each approximation we use shifted inverse square weight function defined as 
$w^{-1}(R) = 1 + {\delta\sigma}^2$, where $\delta\sigma$ is the data 
inaccuracy for the velocity dispersion at radius $R$. 
In Fig.~\ref{fig:obs_data} the solid line presents the approximation 
of the data for the minor axis, and dashed line refers to the data 
for the major axis. The dashed line is drawn to show the difference 
between two profiles, as we do not use the fitting of the major axis data in our analysis.

The ratio $\sigma_{\varphi}^{2}/\sigma_{R}^{2}$ in equation~\eqref{eq:vphi} 
was calculated directly using derivative for the spline 
approximation of $\bar{v}_{\varphi}$. Since we use a smooth spline 
approximation for the rotation curve the derived profile 
$f(R) = \sigma_{\varphi}^2(R)/\sigma_{R}^2(R)$ also stays smoothed 
for all galaxies. Note that values of $f(R)$ usually lie 
between 1 and 0.5, because in the central parts, where the disc rotates solidly,$f(R) = 1$ from equation~\eqref{eq:vphi} and at the 
periphery, where $\bar{v}_{\varphi}$ is almost constant, $f(R)$ is close to 0.5.

If the approximations for the profiles $\bar{v}_{\varphi}$ and 
$\sigma_\mathrm{los, min}$ are fixed ($F(R)$ is also fixed), 
the system of equations~\eqref{eq:final} becomes linear with two 
unknowns $\sigma_{R,0}$ and $\alpha$. The unknown $\sigma_{R,0}$ 
depends on the choice of the point $R_{0}$. We can choose 
any $R_{0}$ within an appropriate range. If we take the point 
$R^{\prime}_{0}$ instead of $R_0$, it will be equivalent to the 
linear substitution of the unknown in the original system of 
equations~\eqref{eq:final} because of the equality~\eqref{eq:f}. 
Thus, the solution of the system does not depend on the choice 
of $R_0$. Here, and after, we assume the value of the radial velocity dispersion at the effective 
bulge radius i.e. $\sigma_{R,0} = \sigma_R(r_\mathrm{e,b})$ as 
an unknown, unless otherwise specified.

Because of noisy data profiles, trying to solve the system of equations~\eqref{eq:final} directly is unreasonable and may led to ambiguous results. That is why we scanned the space of all possible values of unknowns and plot the 
maps of reduced $\chi^2$ values, i.e. weighted sum of squared errors divided by the number of degrees of freedom. We varied $\sigma_{R,0}$ and $\alpha$ in the appropriate range and for every pair of 
$\sigma_{R,0}$ and $\alpha$ we calculated $\chi^2$ for 
the data and expected values predicted by equation~\eqref{eq:final}. It was 
done for both axes data lying after the bulge effective radius 
$r_{e,b}$. From these maps, we can see directly whether there is 
evidence for degeneracy or for the optimal parameters existence. 
We chose the range for parameter $\alpha$ 
as $0.25 \le \alpha \le 1.0$, where the lower boundary is fixed 
at the level just near that given by the local linear criterion 
for the bending instability 
\label{bind}(e.g. \citealp{Rodionov_Sotnikova2013}). 
The upper boundary is generally accepted and has two motivations. 
On the one hand, there are some agent in a galaxy that could 
`transfer' in-plane velocity dispersion into the vertical one 
making a disc thicker. It could be giant molecular clouds (GMCs) 
that are usually thought as a `three-dimensional heating agent'. \citep{Lacey1984} found that when stars are 
scattered by GMCs, a ratio $\alpha \approx 0.8$ is rapidly 
established. On the other hand, in the absence of a third integral of 
motions, the ratio $\sigma_{z}/\sigma_{R}$ to be equal 
to 1 \citep{Binney_Tremaine2008}.
The varying step was equal to $0.01$. For second parameter 
$\sigma_{R,0}$ boundaries were zero as a lowest possible value 
and 500~km/s as an upper limit. The varying step was 0.25~km/s.

\citet{Gerssen_etal1997,Gerssen_etal2000,Shapiro_etal2003,Gerssen_Shapiro2012} used the additional
AD equation for including in dispersion's $chi^2$ also the gas rotation curve (fitted by power function). They obtain a best fit for five-parameters model, while we search in two-dimensional space only, using fixed spline approximations for data profiles, including star rotation curve. Thus proposed method modification is less model-specified and can handle profiles with peculiarities, which are not fitted well by exponential or power law (like NGC~338). However, it has not used additional information and constraint from gas data.

For the obtained values of $\sigma_{R,0}$ and $\alpha$ we can 
retrieve all components of SVE. The profiles of 
$\sigma_\mathrm{los,min}(R)$ and $\sigma_\mathrm{los,maj}(R)$ 
can be directly obtained from equation~\eqref{eq:final}. The stellar velocity 
dispersion profile $\sigma_{R}$ is equal to 
$\sigma_{R,0} \times F(R)$ by definition. The vertical velocity 
dispersion profile $\sigma_{z}(R)$ can be obtained as $\sigma_{R}(R)$ 
multiplied by a factor $\alpha$. Multiplication of 
$\sigma_{R}(R)$ and $f(R)$ from equation~\eqref{eq:f} produces the azimuthal 
dispersion profile $\sigma_{\varphi}(R)$.

\section{Results}

\subsection{NGC~1068} 

In order to check whether our method is working and to 
show the representation of our results for all galaxies, we took 
galaxy NGC~1068 from a different sample \citep{Shapiro_etal2003}. This sample contains another three galaxies, NGC~2460, NGC~2775 and NGC~4030. While NGC~2775 we use in analysis below, other two galaxies we consider insufficient for it because of small number of observational points and short profile's radial region. Selected for test NGC~1068 is Sb type galaxy with the inclination $30 \pm 9$ degrees. The velocity dispersion profiles and the rotation curve lie in 
the $38-90{\arcsec}$ range and the photometric scale for this 
galaxy is equal to $21{\arcsec}$. 
\citet{Shapiro_etal2003} suggested the exponential profiles for 
the radial and vertical velocity dispersions with the same scale 
length. They found the optimal model that minimizes the total 
$\chi^2$ for both $\sigma_\mathrm{los}$ profiles and for the 
AD profile simultaneously. 
In \cite{Shapiro_etal2003} the best fit was reached for 
$\sigma_z / \sigma_R = 0.58 \pm 0.07$ and for 
$\sigma_{R,0}(0) = 213 \pm 20$~km/s.

\begin{figure}
\includegraphics[width=\columnwidth]{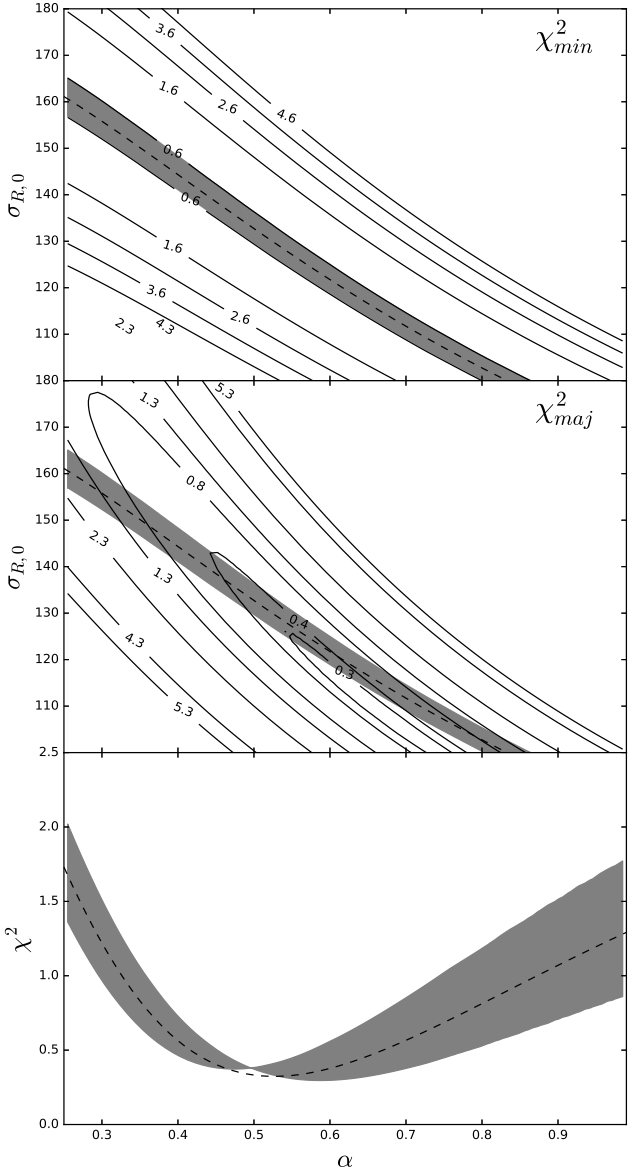}
\caption{Maps of $\chi^2$ error for NGC~1068. In the upper and middle 
plots $\chi^2_{min}$ and $\chi^2_{maj}$ maps are shown for $\sigma_\mathrm{los,min}$ and $\sigma_\mathrm{los,maj}$ data respectively. In both maps solid lines 
represent levels of the same $\chi^2$ and the dashed lines show 
the dependence between $\sigma_{R,0}$ and $\alpha$ (see text for 
details). The grey shaded region corresponds to the values of the 
parameters $\sigma_{R,0}$ and $\alpha$ which minimize $\chi^2_{min}$ 
for the line-of-sight stellar velocity dispersion along the minor 
axis. The bottom plot shows $\chi^2$ values, received for 
$\sigma_\mathrm{los,maj}$ profile when parameters are taken from the 
grey area or along the line relation between $\sigma_{R,0}$ and 
$\alpha$ (dashed line).
}
\label{fig:n1068_map}
\end{figure}

Although, we used the spline fitting instead of exponential functions 
our approximations of the profiles resemble those obtained 
in~\citet{Shapiro_etal2003}. The $\chi^2$-maps for these 
approximations applied for the minor and major axes are presented 
in Fig.~\ref{fig:n1068_map}. The unknown $\sigma_{R,0}$ is the value 
of the stellar radial velocity dispersion at $38.5{\arcsec}$ radii, 
which corresponds to the distance from the galaxy centre to the 
first observational point. To demonstrate the existence of the global 
minimum, we selected the minimal $\chi^2$ value for both unknowns 
($\sigma_{R,0}$ and $\alpha$) for the minor axis with a narrow 
strip around this value. Then we put this strip over the 
$\sigma_\mathrm{los,maj}$ map. In Fig.~\ref{fig:n1068_map} this 
strip is filled by grey and contains all $\chi^2$ values that are 
smaller or equal to the minimal $\chi^2$ plus $10\%$. The minimal reduced $\chi^2$ value 
for $\sigma_\mathrm{los, min}$ is the solution of the first equation 
in equations~\eqref{eq:final}. We drew it as a dotted line. 
Values of $\chi^2$ for the major axis were received for each pair of 
parameters $\sigma_{R,0}$ and $\alpha$ from the grey area and are shown in the bottom subplot in Fig.~\ref{fig:n1068_map}. The dotted 
line is not completely inside the grey area. For example, for 
$\alpha = 0.5,$ two borders of the grey zone intersect the same 
isoline while dotted line has just crossed it and now corresponds 
to a smaller value of $\chi^2$ (the middle subplot in Fig.~\ref{fig:n1068_map}).

The maps have illustrative character and the width of the grey 
strip was chosen arbitrary. We use these maps because they show the 
errors of fitting for both axes independently. In order to determine 
the global minimum correctly, the mutual $\chi^2$-map for both axes 
should be used. It is clear that the $\chi^2$ minimum 
for both axes is reached exactly at the same region where lies 
the minimum for parameters from the grey area in the bottom 
subplot. We plot mutual $\chi^2$-maps for all galaxies in sample except NGC~1167. Such maps are shown in 
Fig.~\ref{fig:n338_pure_chi}--\ref{fig:n4150_map}. 

The constructed maps show that the velocity dispersion profiles are 
not restored individually, as parameters are degenerate. In other words, for any value of $\alpha$ one can find 
a suitable value of $\sigma_{R,0}$ so that an error corresponding 
to the recovered profile will be minimal. For NGC~1068 as well as 
for all other galaxies under consideration, the degeneracy along 
the minor axis is stronger, i.e. the choice of the appropriate values 
of parameters gives the same $\chi^2$ for all $\alpha$. Such a 
behaviour follows directly from the first equation 
in equations~\eqref{eq:final}, where the connection between $\alpha$ and 
$\sigma_{R,0}$ can be explicitly written in the analytical form. 
Nevertheless, it is clear from Fig.~\ref{fig:n1068_map} that a global 
minimum exists. In other words, there are values for $\alpha$ 
and $\sigma_{R,0}$ that minimize errors for both data projections 
simultaneously. The range for the minimal values for $\alpha$ is 
wide enough but still lies inside confidential interval for this 
parameter from \cite{Shapiro_etal2003}. The obtained value for 
the radial velocity dispersion $\sigma_{R,0}$ is consistent with the value determined by \cite{Shapiro_etal2003} after we sample their radial velocity dispersion at $R_0$ (see below).

The result was proven to be stable with Monte-Carlo method using 
10\,000 random realizations of data. For each realization and 
each point along a slit a new ``observable'' was chosen randomly 
from the normal distribution centred at the original value and with 
the dispersion equal to the half of the original uncertainty. 
For this new dataset, new approximations were built and new profiles 
$F(R)$ and $f(R)$ were derived. Then we sorted out all possible 
parameters and found optimal values of $\alpha$ and $\sigma_{R,0}$ 
so that mutual $\chi^2$ for both profiles $\sigma_\mathrm{los, min}$ 
and $\sigma_\mathrm{los, maj}$ would be minimal. Such optimal 
parameters correspond to the global minimum in our maps of $\chi^2$. 
The cloud of 10\,000 optimal parameter pairs was sufficiently compact 
and provides final values of $\alpha$ and $\sigma_{R,0}$ with 
$1\sigma$~error: $\alpha = 0.62 \pm 0.07$, $\sigma_{R,0} = 134 \pm 9$. 
The second parameter $\sigma_{R,0}$ differs from that obtained 
in \citet{Shapiro_etal2003}, because we considered $\sigma_{R,0}$ 
at $38.5\arcsec$ radius. In order to compare the radial velocity 
dispersion values between different models we should correct them for 
distance to the centre. For this galaxy the kinematic exponential scale length 
obtained in \citet{Shapiro_etal2003} is equal to $72\arcsec$, hence 
a correction factor is equal to $\exp(-38.5/72.)$ and produces the 
final value of $\sigma_{R,0}$ at $R = 38.5\arcsec$ to be equal 
to 125~km/s. This final value is very close to that obtained in our 
analysis.

\subsection{Examined galaxies}

\begin{figure}
\includegraphics[width=\columnwidth]{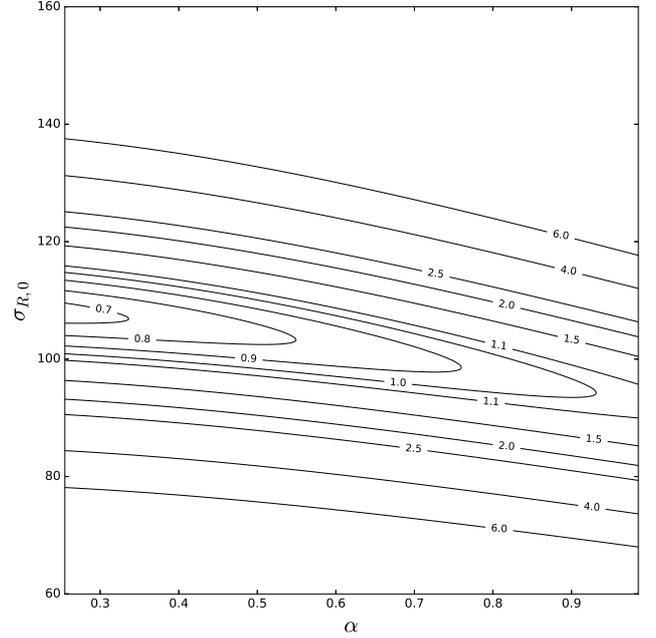}
\caption{The map of mutual $\chi^2$ for the major and minor axes 
for the galaxy NGC~338. The lines show levels of the same $\chi^2$ 
values with the level value written in the plot near each line.}
\label{fig:n338_pure_chi}
\end{figure}

\begin{figure}
\includegraphics[width=\columnwidth]{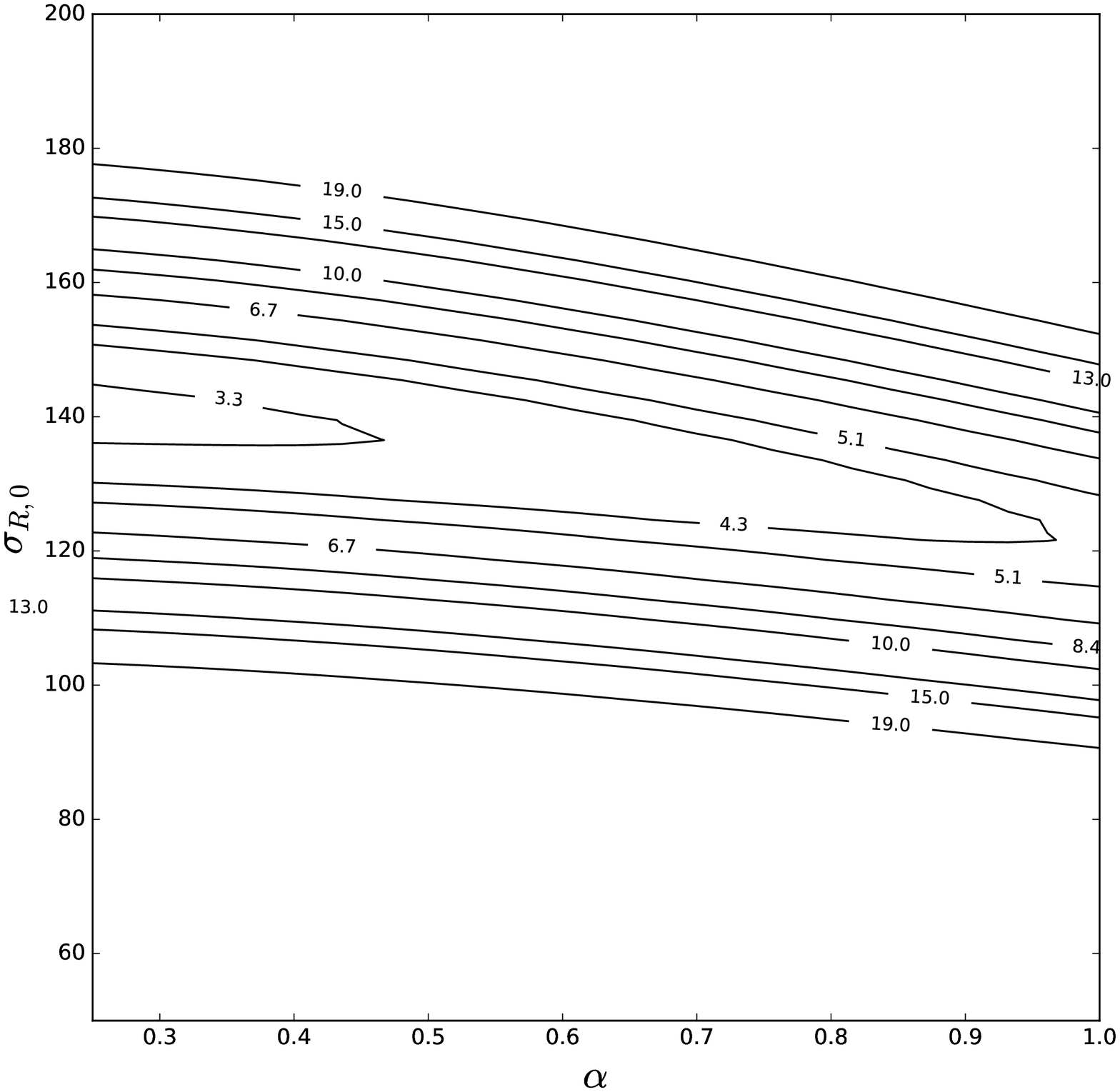}
\caption{Map of mutual $\chi^2$ for NGC~3245 (similar to Fig.~\ref{fig:n338_pure_chi}).}
\label{fig:n3245_map}
\end{figure}

\begin{figure}
\includegraphics[width=\columnwidth]{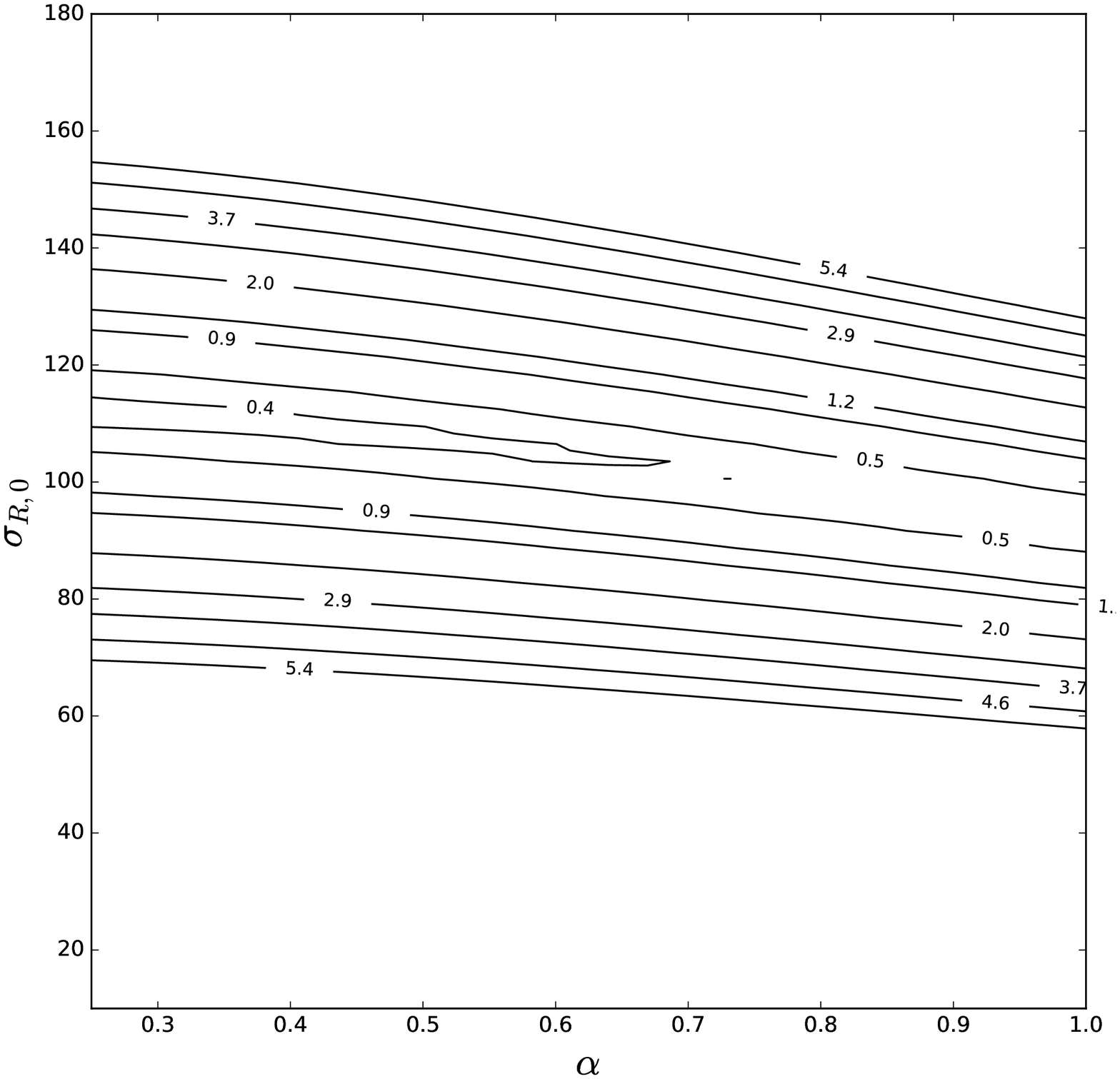}
\caption{Map of mutual $\chi^2$ for NGC~4150 (similar to Fig.~\ref{fig:n338_pure_chi}).}
\label{fig:n4150_map}
\end{figure}

\begin{figure}
\includegraphics[width=\columnwidth]{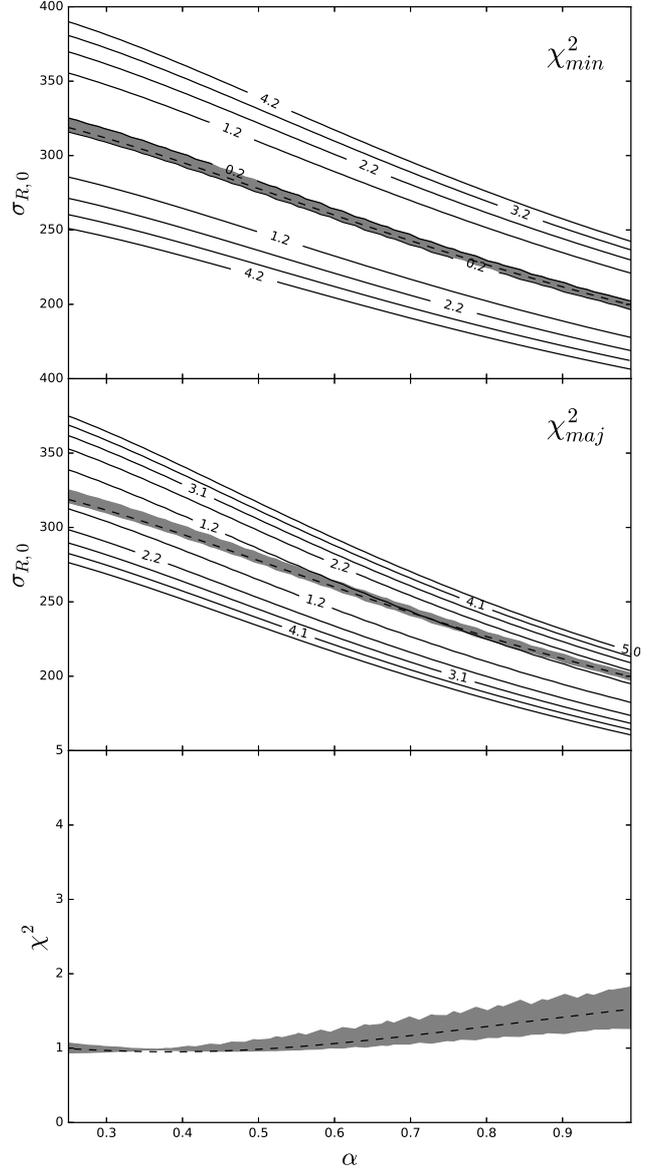}
\caption{Maps of $\chi^2$ for NGC~1167. Notations similar to those in 
Fig.~\ref{fig:n1068_map}.}
\label{fig:n1167_map}
\end{figure}

We obtained similar maps and performed Monte-Carlo checking as 
before for all galaxies from the Table~\ref{table:main_parameters}. Obtained results 
show no evidence of global minimum existence, except one galaxy. 
All maps for examined galaxies are drawn in 
Fig.~\ref{fig:n338_pure_chi}--\ref{fig:n1167_map} in the same manner as 
for NGC~1068. The exception is NGC~1167 that has the lowest inclination 
angle among the galaxies in the Table~\ref{table:main_parameters}. For this galaxy, 
our analysis gave the minimum for $\sigma_z/\sigma_R$ near the 
lowest boundary for $\alpha = 0.3$. 
However, this minimum was reached not only in a single point. Values of 
$\chi^2$ stay almost constant in the wide range from lowest 
$\alpha$ (even for $\alpha < 0.3$) till $\alpha \approx 0.5$, i.e. 
minimum is very vast. Monte-Carlo modeling does not produce any 
preferable value of $\alpha$ from $\alpha < 0.5$ range. Thus, the case 
of NGC~1167 is not simple and this galaxy will be examined separately 
in the next section.

For other three galaxies, NGC~338, NGC~3245 and NGC~4150 
$\chi^2$ monotonically increases with growing 
$\alpha$. Hence, the lowest $\chi^2$ may be found for the lowest 
$\alpha$ or, in other words, at the left boundary of $\alpha$, 
as it shown in the bottom plots of the appropriate map. 
If we consider maps for both axes independently they will 
demonstrate a degeneracy trend in the same way as for NGC~1068. 
The difference in $\chi^2$ values between these three galaxies may be 
explained by galaxies' parameters dissimilarity, 
namely, by different typical uncertainties, different number of data 
points and so on. For example, 
NGC~3245 has small relative errors and a large gap between 
$\sigma_\mathrm{los, min}$ and $\sigma_\mathrm{los, maj}$ profiles 
that leading to large $\chi^2$ values.

Small optimal values of $\alpha$ are likely not to be naturally occuring and 
appear as artefacts in numerical models due to the limited range of 
$\alpha$. These small values were confirmed by Monte Carlo simulations 
that produce $\alpha$ near to the left border in most realizations. 
As mentioned in Section~4, $\alpha$ less than 0.3 
seems to be unrealistic. Thus, obtained results make questionable our main 
assumption about small variation of the ratio $\sigma_z/\sigma_R$. Another possible source of uncertainties is 
observational data quality. However, presented list of galaxies 
is not uniform by many parameters and it is unlikely 
that data quality may cause similar results especially 
with additional statistical support. Small values of $\alpha$ 
and difficulties in its extraction from data are due to 
the small contribution of $\sigma_z$ in the observed 
line-of-sight velocity dispersion profiles when a galaxy has 
a large inclination. We discuss this in detail in the next section.

\section{Discussion}

The results for three galaxies in the Table~\ref{table:main_parameters} differ from the 
result for NGC~1068. In these galaxies, a strong degeneracy of 
parameters is observed and all formal optimal values of $\alpha$ 
are as small as possible in the physically meaningful range. 
Even for NGC~1167, for which a real global minimum of $\chi^2$ 
may be found ($\alpha\approx 0.3$), more accurate analysis should 
be done. 
Received results may relate to the impact of a significant inclination 
on observational data or may be caused by incorrectness of the main 
assumption of small changes of $\alpha$ across kinematic profiles. 
Both possibilities will be examined below. We also discuss 
whether the boundary values of $\alpha$ may be considered as physical. 
In last part of this section, we estimate 
the possible range for the radial velocity dispersion values 
and analyse its consistency with the range constrained by other methods. 

\begin{figure}
\includegraphics[width=\columnwidth]{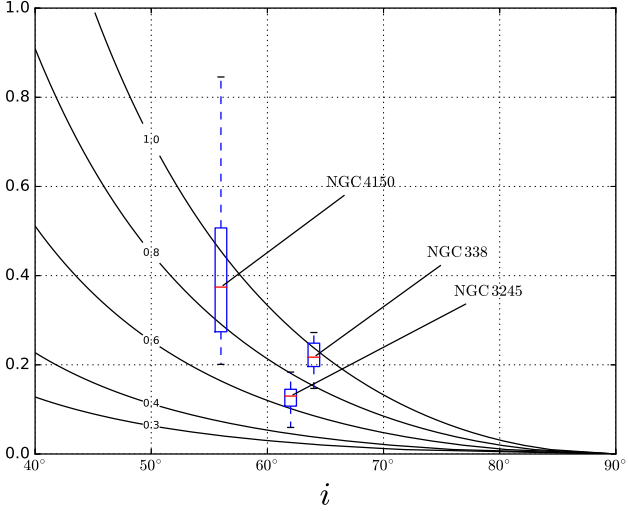}
\caption{Inclination impact comparison with observational uncertainties. Solid lines show $\alpha^2\mathrm{ctg}^2i$ for different $\alpha$, which correspond to vertical velocity dispersion term contribution in second equation in equations~\eqref{eq:slos}. Distributions of $\sigma^2_\mathrm{los, min}$ relative errors for three annotated galaxies are represented by boxplots, which demonstrate maximum, minimum, median, first and third quartiles of distribution. Every boxplot have the same width and x-coordinate equal to corresponding galaxy's inclination.}
\label{fig:incl_impact}
\end{figure}

The inclination impact may be evaluated from equations~\eqref{eq:slos}. 
According to these equations, the vertical velocity 
dispersion contributes to the observational data a little when the galaxy inclination is large. 
Indeed, it is seen from second equation that the stellar 
line-of-sight velocity dispersion for the minor axis consists of 
two terms, $\sigma^2_R\sin^2i$ and $\sigma^2_z\cos^2i$, that 
contribute in the resulting sum as 1:$\alpha^2\mathrm{ctg}^2i$. 
For the inclination angles $i > 60\degr$ this leads to the vertical 
velocity dispersion term contribution not greater than one-third. 
This value should be compared with the observational uncertainties.
As shown on Fig.~\ref{fig:incl_impact}, in the case of NGC~338 and NGC~4150 the relative 
$\sigma^2_\mathrm{los, min}$ errors are large and thus for 
every $\alpha \le 0.7$ the contribution of $\sigma_z$ is comparable 
with observational uncertainties and cannot be correctly extracted. 
Even for NGC~3245, which has the smallest relative errors in the sample, for $\alpha \le 0.7$ the vertical velocity dispersion cannot be distinguished from the error for majority of observational points and thus it is also 
insufficient for the correct reconstruction of SVE. Note that it appears 
unlikely for our data that it is possible to obtain larger vertical 
to radial velocity dispersions ratio, because we 
analyse extended profiles and in the absence of GMCs in the outer 
part of a galaxy there is only merging that can lead to the
high vertical disc heating till distant radii 2-2.5~$h$. Heating by 
merger may be true for NGC~338, but NGC~3245 is isolated and 
NGC~4150 has poor environment. 

Thus for galaxies with the inclination $i > 60\degr$ and 
relative data errors similar to ours it is hard or nearly 
impossible to reconstruct $\alpha$ values. Second difficulty is that 
the equation~\eqref{eq:los_comm} is valid only for a thin disc and 
does not include the line-of-sight integration of the DF for galaxies with high inclination. 
Perhaps, this is the reason why we obtain the formal minimal 
$\chi^2$ values for the smallest values of $\alpha$ for NGC~338, 
NGC~3245 and NGC~4150.

\subsection{The $\alpha$ solution on the edge}

There exists one simple reason why obtained values of $\alpha$ 
on the edge of the variation range are not realistic. It can be seen 
from the $\chi^2$-maps that for three galaxies NGC~338, NGC~3245 
and NGC~4150 $\chi^2$ continues to decrease outside the left border. 
If we did not know an additional constraint for the ratio 
$\sigma_z/\sigma_R$, we could derive unphysical $\alpha$ values 
from the analysis. 
Optimal $\alpha$ values close to their variation boundaries were 
received in papers by other authors too. 
For example, the optimal $\alpha$ values were found both equal to the maximal 
possible value $\alpha = 1$ (\citealp{Shapiro_etal2003}, NGC~2775) 
and near the left boundary $\alpha = 0.25$ for NGC~2280 and 
$\alpha = 0.29$ for NGC~3810 \citep{Gerssen_Shapiro2012}. For these three galaxies only NGC~2775 has data long enough for analysis. We analyse it below in order to determine whether such optimal values on the edge were realistic or not.

NGC~2775 is a galaxy of Sa or Sab morphological type and has the inclination angle 
around $40\degr$. The velocity dispersion profiles in 
\citet{Shapiro_etal2003} are extended in the wide radial region $31-62{\arcsec}$. They mention that there are 
differences in bulge contribution evaluations between papers and it 
is possible that the disc does not dominate in the examined range. 
However, they believe that the bulge influence becomes negligible 
for distances greater than $30-40\arcsec$ from the galaxy centre. 
The best model fit for $\alpha$ was $1.02 \pm 0.11$. 

\begin{figure}
\includegraphics[width=\columnwidth]{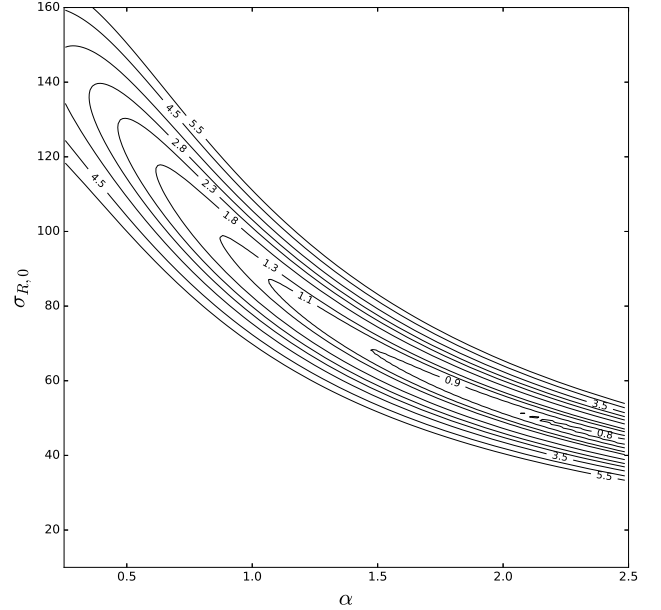}
\caption{Map of $\chi^2$ for NGC~2275 (similar to Fig.~\ref{fig:n338_pure_chi}).}
\label{fig:n2775_map}
\end{figure}

Our best approximation of the profile $\sigma_\mathrm{los,min}$ is 
close to the best fit in \citet{Shapiro_etal2003}. It may be 
explained by the fact that observational points lie not far from 
exponential fit and splines follow the same trend. 
Thus, it is expected that when we try to modify the method by
using exponential profile parametrization and apply it to NGC~2775, 
we obtain quite the same results. The rotation curve changed a little
at radii under discussion and also was fitted 
quite similar to the approximation in the original paper. 
The $\chi^2$-maps are shown in Fig.~\ref{fig:n2775_map}. 
In order to be confident that $\alpha$ being close to the 
right boundary is not a true minimum, we extended the range of 
$\alpha$ and plotted the maps for greater $\alpha$ up to 
$\alpha = 2.5$. 
Fig.~\ref{fig:n2775_map} shows that minimal $\chi^2$ formally reaches 
for $\alpha = 1$ and $\sigma_{R,0} \approx 90$~km/s, which are close 
to the values from \citet{Shapiro_etal2003}. 
However, $\alpha = 1$ is not a global minimum, since $\chi^2$ 
continues to decrease drastically and reaches notably lower 
values for larger $\alpha$. We check the stability of this 
result with 10\,000 Monte Carlo realizations of the data. 
For all realizations we found the minimal $\chi^2$ value for $\alpha$ 
to be close to 2.5. 

Hence, the galaxy NGC~2775 demonstrates that the optimal $\alpha$ 
being close to the limit of possible physical values is not a feature 
of observational data or selected approximations. 
Despite that we derived the same optimal parameter for $\alpha \le 1$ (as in \citealp{Shapiro_etal2003}), we 
showed that the minimal $\chi^2$ value lies far from constrained 
boundary. In other words, the solution $\alpha = 1$ is not real 
and seems rather to be a numerical peculiarity of the model conditions. 
We also tried to perform the analysis for NGC~2280 from 
\citet{Gerssen_Shapiro2012} that has small $\alpha$, but failed to reproduce the result as
 we obtained a high value of $\alpha$.

\subsection{Non-constant $\alpha$ model for NGC~1167}

The galaxy NGC~1167 is inclined at $30\degr$ and thus contribution of $\sigma_z$ in $\sigma_\mathrm{los,min}$ become significant. In addition, the galaxy demonstrates a real global 
minimum for $\alpha$ inside the possible range. 
However, the optimal value $\alpha \approx 0.3$ lies near the 
boundary where disc stays stable against bending instability 
(see Section~4). Since there are many dynamical mechanisms 
for the disc heating in the vertical direction, this minimal value 
seems to be questionable. The only substantial assumption in our 
model, which may affect the position of global minimum, is 
$\alpha = \mathrm{const}$ throughout the disc. This may be wrong 
for extended profiles. 
Therefore we tried to derive $\alpha$ for two smaller separate 
pieces of the profiles and compared them. 
We make such an analysis for NGC~1167 and discuss derived results 
below.

The profile $\sigma_\mathrm{los, min}$ demonstrates a bend at about 
$30\arcsec$ (Fig.~\ref{fig:obs_data}). We took this point to split up 
the profile into two parts and found the best values of $\alpha$ before 
and after the bend independently. We examined the profile 
at the ranges from the bulge effective radius $r_\mathrm{e,b}$ 
till $30\arcsec$ and from $30\arcsec$ till the end of the 
observational data. These pieces of the profile have roughly 
equal lengths. We approximated them by exponential law because 
they are well described by such parametrization. Then for 
every part the $\chi^2$-maps were constructed and the optimal values of 
$\alpha$ were found at the same manner as before. Obtained 
values of $\alpha$ were tested using the Monte Carlo method 
with 10\,000 realizations of the data for every part.

\begin{figure}
\includegraphics[width=\columnwidth]{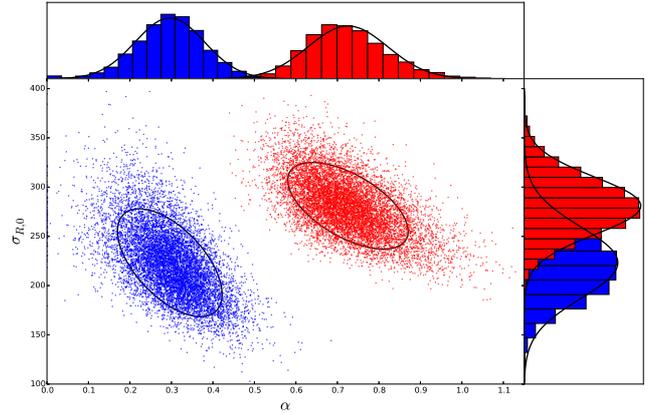}
\caption{The results of Monte-Carlo simulation for galaxy NGC~1167 
for the profile $\sigma_\mathrm{los, min}$ divided in two parts. 
Each point shows optimal values of $\alpha$ and $\sigma_{R, 0}$ 
for a particular realization of data that minimizes mutual $\chi^2$. 
A cloud with red dots and greater $\alpha$ corresponds to the 
inner part of the profile while the blue dots with smaller 
$\alpha$ correspond to the outer part. 
A solid ellipse for each cloud demonstrates the level of one 
standard deviation from the centre of the two-dimensional normal 
distribution. The upper and right subplots show histograms of each 
parameter with the solid line that follows the best Gaussian 
approximation.}
\label{fig:MC}
\end{figure}

The results of simulations are shown in Fig.~\ref{fig:MC}. It is seen that the clouds of Monte Carlo realizations for 
different parts are separated clearly from each other. These clouds 
are relatively compact and well described by a two-dimensional normal 
distribution for $\sigma_{R,0}$ and $\alpha$. For the inner part 
we found $\alpha = 0.72\pm0.09$ with $1\sigma$ error. For 
the outer part ($30-60\arcsec$) $\alpha = 0.30\pm0.08$. 
Second value of $\alpha$ lies close to the limit of the physical 
parameter range but is probably real. In isolated galaxies, 
in the absence of a separate agent causing out-plane scattering 
(GMCs, nearby external galaxies), the initial bending 
instability, which develops and decays within the first Gyr of 
the disc evolution, heats outer parts of the disc up to the level
corresponding to the linear criterion of the bending 
instability $\alpha \approx 0.30$ 
\citep{Rodionov_Sotnikova2013}.

\begin{figure}
\includegraphics[width=\columnwidth]{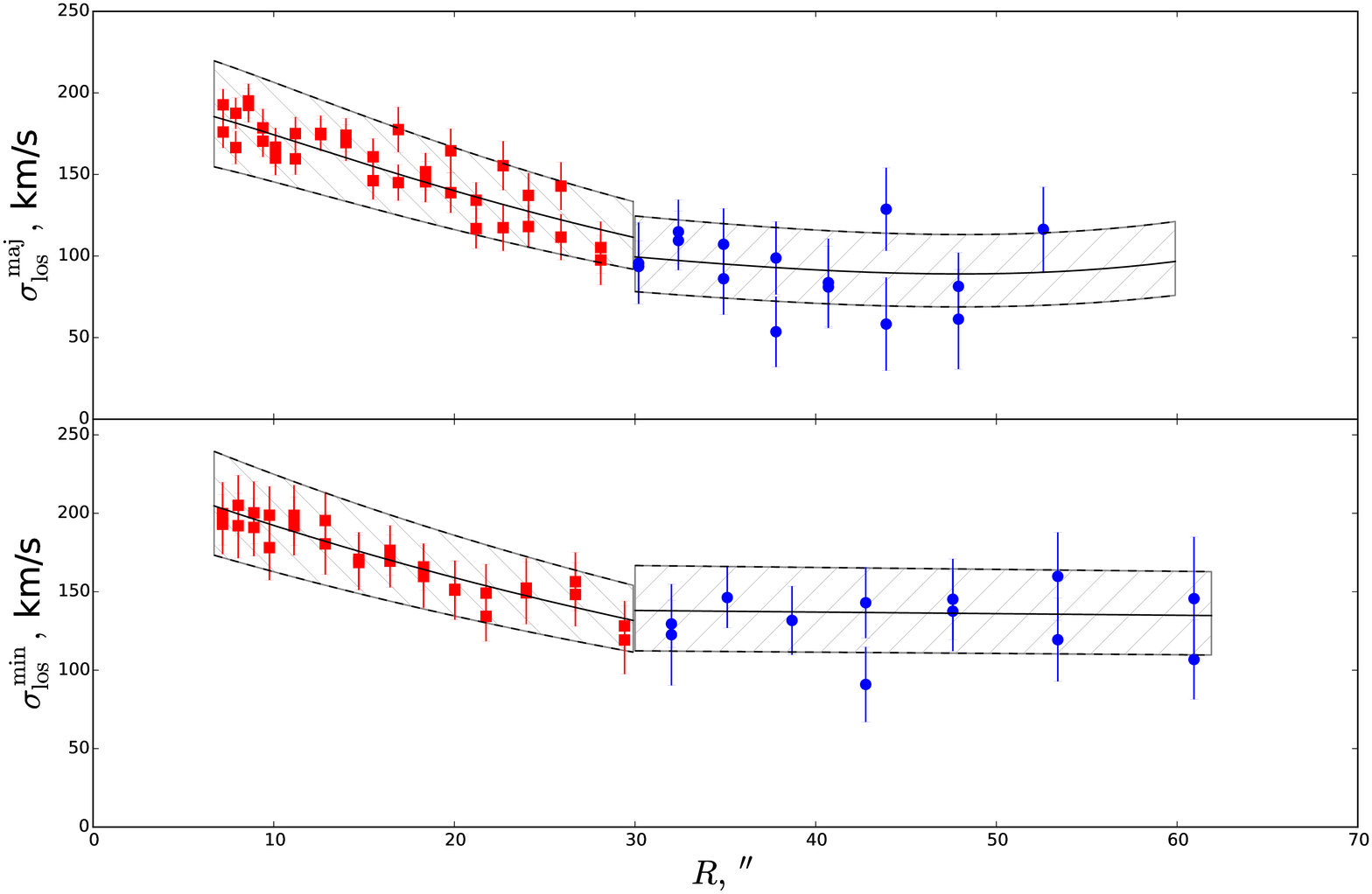}
\caption{Observed and reconstructed profiles of the line-of-sight 
velocity dispersions for NGC~1167 in the case of divided profile. 
The upper plot shows the profile of the line-of-sight velocity 
dispersions along the major axis $\sigma_\mathrm{los,maj}(R)$ 
and the bottom plot shows the profile along the minor axis 
$\sigma_\mathrm{los,min}(R)$. 
For both plots, filled symbols indicate observational 
data (red squares for the inner part and blue circles 
for the outer one). The solid line 
represents a reconstructed velocity dispersion profile 
with parameters equal to the mean values of bivariate normal 
distributions obtained by Monte Carlo method (see Fig.~\ref{fig:MC}). 
Shaded area corresponds to all possible model profiles 
$\sigma_\mathrm{los,maj}$ and $\sigma_\mathrm{los,min}$ 
reconstructed for parameters $\alpha$ and $\sigma_{R,0}$ which 
differ from its mean values no more than one standard deviation.}
\label{fig:n2775_slos}
\end{figure}

The decrease in $\alpha$ with rising distance from the centre is consistent with 
the results of the $N$-body simulations by \citet{Zasov_etal2008}.
It is also remarkable that our value of $\alpha$ in the inner part 
is found to be equal to the one obtained in~\citet{Zasov_etal2008} using $N$-body simulations (see fig.~8 in the cited paper). 
For both parts we reconstructed the original line-of-sight velocity 
dispersion profiles in the manner specified above. These profiles 
with the acceptable deviations are shown in Fig.~\ref{fig:n2775_slos}. Both 
profiles are consistent with observational data. 
This fact gives a strong support for the obtained 
results. Notice that the solid line, which shows the inner reconstructed 
profile for $\sigma_\mathrm{los,min}$ (the bottom plot in 
Fig.~\ref{fig:n2775_slos}), is almost identical 
to exponential approximations used in the analysis.

It can be noted that optimal $\alpha$ determined for the whole profile 
is not equal to the average between values of optimal $\alpha$ 
obtained for two separated parts of the profile and is close 
to the lowest one. There are two reasons for such behaviour. First, the global minimum for $\alpha \approx 0.3$ is 
very vast and the change of $\chi^2$ is almost not noticeable up to 
the value of $\alpha \approx 0.5$, which is close enough to the 
average value. 
The second reason is that the observed profiles of 
$\sigma_\mathrm{los,min}$ and $\sigma_\mathrm{los,maj}$ are 
significantly diverge at distant regions. 
If we constrain $\alpha$ between $0.3$--$0.5$ this results in small $\chi^2$ values. 
Increasing values of $\alpha$ will produce the profile 
$\sigma_\mathrm{los,maj}$ with more and more deviation from the 
observational data at large distances from the centre that results in the $\chi^2$
increase. At the same time, the central part of the major axis 
profile is approximated rather well for small values of $\alpha$, 
only slightly increasing the total error. These two reasons explain 
why $\alpha \approx 0.3$ for the whole profile.

Thus, for NGC~1167, the initial assumption of a constant value of 
$\alpha$ along the whole profile is less physically motivated. 
If we reject this assumption we can reconstruct SVE 
over the entire radial region and find using Monte Carlo simulations the obtained inner and outer values of $\alpha$ and $\sigma_{R,0}$ to be stable. However, for other galaxies, we did not perform such analysis because for them inclination impact cannot be omitted even for splitted profile.

\subsection{AD and constraints on $\sigma_{R,0}$ for NGC~338} 

Although we failed to determine the SVE and find the optimal $\alpha$ 
for the highly inclined galaxies, we can get a constraint to the 
second parameter $\sigma_{R,0}$. This constraint follows from the 
$\chi^2$-maps (Fig.~\ref{fig:n338_pure_chi}--\ref{fig:n4150_map}). 
It is seen, that the factor $\sigma_{R,0}$ for galaxies with large 
inclinations changes a little in a wide range of $\alpha$ values. 
Such a behaviour infers from the first equation in equations~\eqref{eq:final} 
where the relation between $\sigma_R$ and $\sigma_\mathrm{los,min}$ 
is defined by the value in brackets, depending on two parameters 
$\alpha$ and $i$. 
The ratio $\alpha = \sigma_z/\sigma_R$ can formally vary between 
0 and 1. Then, the value in brackets varies from $\sin^2 i$ 
to 1, respectively. Therefore, for inclinations about $60\degr,$ the 
corresponding value of $\sigma_R$ at a given point changes slightly and remains 
within 1--1.25 values of $\sigma_\mathrm{los,min}$. 
This constraint does not take into account the data along the 
major axis, but can be checked independently using the gas 
rotation curve $v_\mathrm{c}$. We performed such a test 
for the galaxy NGC~338.

NGC~338 is the only one in the Table~\ref{table:main_parameters} for which the gas rotation 
data are good enough. These data were obtained from the analysis 
of $\mathrm{H}_\beta$ and $\mathrm{[OIII]}$ emission lines 
and are described in detail in \citet{Zasov_etal2012}.

We used the equation~\eqref{eq:AsDr} with known photometry and 
a set of $\sigma_{R, 0}$ values in a fixed range to determine 
the value of AD. 
This value should then be added to the average azimuthal velocity of 
stars to obtain the model profile $v_\mathrm{c}$ that should be 
compared with the observed rotation curve. 
After going through all the possible values of 
$\sigma_{R, 0}$ and calculating for them $\chi^2$ error we 
managed to find the optimal rotation curve that minimizes the difference between the gas data and the model. It should be noticed 
that for large inclination we need to account for the line-of-sight 
integration of the DF when determining rotation 
velocities. We did not make any reduction of data because for 
$i<60\degr$ the distortion of the rotation curve for $R>h$ is 
minimal \citep{Stepanova_Volkov2013}.

Let us specify how to calculate the expression in brackets in 
the equation~\eqref{eq:AsDr}. The ratio 
${\displaystyle \frac{\sigma_{\varphi}^{2}}{\sigma_{R}^{2}}}$ is 
known and equal to $f$. 
To calculate the logarithmic derivative 
${\displaystyle \frac{\partial\ln\Sigma_\mathrm{s}}{\partial\ln R}}$ 
we assume that the mass-to-light ratio is constant along the disc. 
If so, instead of the surface density $\Sigma_\mathrm{s}$ 
we can use the stellar surface brightness of the disc measured 
in the bands tracing the old stellar population. 
For an exponential profile of brightness, this means that 
the logarithmic derivative should be replaced by 
$-{\displaystyle \frac{R}{h}}$, where $h$ denotes the disc 
scale length. 
The required photometry in the $I$ band was taken from 
\citet{Noordermeer_vanderHulst2007}, where the disc scale length 
was found to be equal to $12.9\arcsec$. 
Finally, the last logarithmic derivative 
${\displaystyle \frac{\partial\ln\sigma_{R}^{2}}{\partial\ln R}}$ 
can be numerically found from the fit of 
$\sigma_\mathrm{los,min}(R)$ profile because 
$\sigma_\mathrm{los,min}(R)$ is proportional to $\sigma_R(R)$ 
with the constant of proportionality $\sin^2i + \alpha^2\cos^2 i$. 

\begin{figure}
\includegraphics[width=\columnwidth]{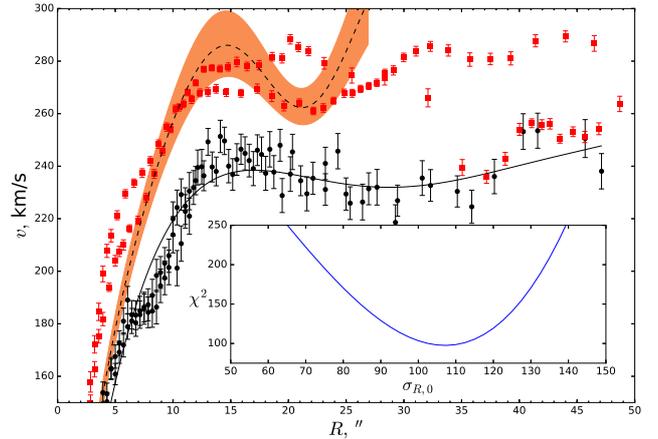}
\caption{Reconstructed gas rotation curve of NGC~338. 
Filled black circles and solid line denote observational points 
and the best-fitting approximation, respectively. 
Red squares shows ionized gas rotation velocities, 
the so-called cold rotation curve. 
The subplot shows $\chi^2$ of gas model due to the AD 
prediction for different values of $\sigma_{R,0}$. 
Gas rotation curve for $\sigma_{R,0} = 107$~km/s corresponding 
to the minimum of $\chi^2$ error is shown by the dashed line. 
The shaded area corresponds to the model gas rotation curves 
reconstructed by using values of $\sigma_{R,0}$ that differ 
by no more than 30~km/s from the optimal value.}
\label{fig:ad}
\end{figure}

Results for NGC~338 are shown in Fig.~\ref{fig:ad}. 
The figure demonstrates the gas and the stellar 
rotation curves, the model curve and the errors depending 
on the model parameter $\sigma_{R, 0}$. The $\chi^2$ curve 
demonstrates a strict global minimum at about 107~km/s. 
The optimal model for $v_\mathrm{c}$ is shown by dashed line 
and the shaded area corresponds to models with $\sigma_{R, 0}$ values 
differing from the optimal ones by less than 30~km/s. 
Notice that for $R \ge 27\arcsec$ the derivative of 
$\sigma_\mathrm{los, min}$ produces large uncertainties 
because of a small number of points in this area and the 
AD is poorly defined. The gas rotation curve 
is approximated quite well within the radial region where model is good 
except areas close to the centre of the galaxy. 

For NGC~338, the value of 
$\sigma_\mathrm{los,min}(r_\mathrm{e,b})$ is close to 100~km/s and hence the range of possible $\sigma_{R, 0}$ values vary
from 100 till 125~km/s as stated above. These constraints agree 
well with the values obtained from the AD analysis. 
Another less significant evidence comes from the mutual $\chi^2$-map, 
calculated for both axes simultaneously in the way described in 
Section~4. This map is shown in Fig.~\ref{fig:n338_pure_chi} and 
demonstrates that the best fit for two observational profiles 
$\sigma_\mathrm{los,maj}$ and $\sigma_\mathrm{los,min}$ 
is obtained for values of $\sigma_{R, 0}$ from the same range 
as above.

Thus, for galaxies with a large inclination, we can obtain 
constraints on the stellar radial velocity dispersion. These constraints produce a narrow range of possible values. Example of NGC~338 shows that this 
range is consistent with the AD arguments. 
The value of $\sigma_{R}$ with the 
rotation curve of the galaxy and the mass model of its disc 
are sufficient to find the value of the parameter $Q$ and 
determine the dynamical status of the stellar disc.

\section{Summary}

Dynamical models of galaxies require knowledge of the spatial stellar 
velocities distribution or at least of three components of SVE.
The velocity dispersion in disc galaxies offers a clue for 
determining the dynamical status of a disc, its stability, dynamical 
history and relaxation processes. 
Unfortunately, velocities of stars in external galaxies 
are measured only in the projection on the line of sight. 
Full three-dimensional picture of the velocity distribution could be
restored from the stellar spectral data indirectly. 
To reconstruct it several methods were proposed: 
via the exponential parametrization of the velocity dispersion 
profiles and subsequent data fitting along the major and minor axes 
\citep{Gerssen_etal1997,Gerssen_etal2000,Shapiro_etal2003,Gerssen_Shapiro2012}; 
by apply the AD equation 
\citep{Noordermeer_etal2008,Silchenko_etal2011}; 
through building a set of $N$-body marginally stable discs to choose the model that approximates spectral data in the best way \citep{Zasov_etal2008}.

In this paper, we present a modification that does not employ a strict 
parametrization of the velocity dispersion profiles and requires only one additional assumption about the 
ratio $\sigma_z/\sigma_R = \alpha$, namely $\alpha$ remains constant 
along the whole profile or along its parts. This assumption allows us to 
introduce the proportionality between $\sigma_\mathrm{los, min}$ 
and the velocity dispersion in the radial direction $\sigma_R$. 
Using this relationship, we can find the optimal value 
of $\alpha$ and the radial velocity dispersion normalization parameter 
$\sigma_{R,0}$ by applying error minimization procedure simultaneously 
to the line-of-sight stellar velocity dispersion profiles 
along the major and minor axes of the galaxy which were pre-fitted. 

Our approach has been successfully tested on the galaxy NGC~1068 and 
showed the results similar to those obtained earlier by 
\citet{Shapiro_etal2003}. For further analysis, we chose three lenticular S0 galaxies NGC~1167, NGC~3245, NGC~4150 and one Sab galaxy NGC~338 with known rotation curves of stars and gas as well 
as the stellar velocity dispersion profiles.

Our results for these galaxies demonstrate a strong degeneracy trend 
for both major and minor axes. The simultaneous data consideration 
along two axes does not allow us to find optimal values of 
parameters in the physically meaningful range for 
galaxies with large inclinations (NGC~338, NGC~3245 and NGC~4150). In 
all cases, there is a gradual growth of $\chi^2$ with $\alpha$ 
increasing. For the galaxy at a moderate inclination NGC~1167, 
as well as for the galaxy NGC~1068, we managed to find the minimum 
of $\chi^2$. The results were verified by Monte Carlo method.

It should be noted that the best way to restore the SVE components 
for NGC~1167 galaxy is to reject the assumption 
$\sigma_z/\sigma_R = \mathrm{const}$ throughout the whole profile 
and to divide the profile into the inner and outer parts, 
along which the main assumption still holds. 
For each of these parts we found the global minimum of $\chi^2$. 
The optimal values of $\alpha$ were found to be equal to 0.7 
for the inner part and around 0.3 for the outer part. 
The decrease in $\alpha$ with rising distance from the centre is 
consistent with the results of the $N$-body simulations for this 
galaxy implemented by \citet{Zasov_etal2008}. Other numerical 
simulations also support such a result 
(see, e.g. \citealp{Minchev_etal2012}). Thus, for the first time, 
the gradient in the meridional shape of the SVE was measured from the 
the spectral data.
Stability of the results was verified using the Monte Carlo 
simulation. Recovered line-of-sight profiles of the stellar velocity 
dispersions for the major and minor axes showed good agreement with 
the observational data. Also, the resulting values of $\alpha$ are 
consistent with the disc dynamical heating theory. NGC~1167 has a 
moderate amount of gas in the molecular form 
($M_\mathrm{H_2} = 3.3 \times 10^8 M_{\sun}$) 
within 7.8~kpc, or 25\arcsec \citep{OSullivan_etal2015}. 
Molecular gas is usually thought as a 'three-dimensional 
scattering agent' that increases the vertical velocity dispersion 
above the level of the bending instability. At the same time, the 
outer parts of a disc may remain 'cold' just at the level 
$\sigma_z/\sigma_R \approx 0.3$ \citep{Rodionov_Sotnikova2013}.

For galaxies with high inclinations SVE was not reconstructed 
because contribution of the vertical velocity dispersion 
component in the line-of-sight velocity dispersion data is comparable with 
observational uncertainties for such galaxies and thus cannot be extracted from the data in practice. 
As a result, the formal optimal value of $\alpha$ falls on one 
of the boundaries of the range of $\alpha$ (either $\alpha$ 
minimal or equal to 1). 
Thus, the result for the galaxy NGC~2775 from the paper 
\citet{Gerssen_Shapiro2012}, where $\alpha \approx 1$ was obtained, 
can be considered as a formal solution.

Despite the fact that the impact of a high inclination does not 
allow us to restore correctly all velocity dispersion components, 
for these galaxies we can constrain the possible values of the radial 
velocity dispersion $\sigma_R$ by a sufficiently narrow range. 
For example, for NGC~338, for which there are good enough gas 
rotation curve data, these constraints were verified with the help 
of the AD equation. Obtained values were in good agreement 
with each other and with the values from the $\chi^2$-maps.

\section*{Acknowledgements}
We are grateful to Alexei Moiseev and Ivan Katkov for providing
spectral data. We thank Saint Petersburg State University research
grant 6.38.335.2016 (AAM) and grant of the Russian Foundation
for Basic Research number 14-02-810 (NYS) for partial support.
We thank the anonymous referee for his/her thorough review and
highly appreciate the comments and suggestions that significantly
contributed to improving the quality of the article.

This research makes use of the NASA/IPAC Extragalactic
Database (NED) which is operated by the Jet Propulsion Laboratory,
California Institute of Technology, under contract with the
National Aeronautics and Space Administration, and the LEDA
database (\href{url}{http://leda.univ-lyon1.fr}).


\bibliographystyle{mnras}
\bibliography{art}

\end{document}